\documentclass[useAMS,usenatbib]{mn2e}
\usepackage{lscape,epsfig}
\usepackage{graphicx}
\usepackage[margin=5pt,font=small,labelfont=bf,
            justification=raggedright,singlelinecheck=false]{caption}
\title[Absolute parameters of AE For]
 {Absolute parameters of AE For -- a highly active detached binary of late
  K type\thanks{Based in part on data obtained at the Las Campanas Observatory.}
 }
\author [M. Rozyczka et al.]
 {M. Rozyczka,$^{1}$\thanks{E-mail: mnr@camk.edu.pl}
  P. Pietrukowicz$,^{2}$
  J. Kaluzny,$^{1}$
  W. Pych$^{1},$
  R. Angeloni$^{3,4}$
  \newauthor
  and I. D\'ek\'any$^{4,5}$
  \\
  $^{1}$Nicolaus Copernicus Astronomical Center, Bartycka 18, 00-716 Warszawa, Poland\\
  $^{2}$Warsaw University Observatory, Al. Ujazdowskie 4, 00-478 Warszawa, Poland\\
  $^{3}$Dept. of Electrical Engineering, Center for Astro-Engineering, Pontificia Universidad
        Cat\'olica de Chile, \\ \hspace{1.7 mm}Av. Vicu\~na Mackenna 4860, Macul, Santiago, Chile\\
  $^{4}$Dept. de Astronom\'{\i}a y Astrof\'{\i}sica, Pontificia Universidad
        Cat\'olica de Chile, Av. Vicu\~na Mackenna 4860, Macul, Santiago, Chile\\
  $^{5}$The Milky Way Millennium Nucleus, Av. Vicu\~na Mackenna 4860, Macul, Santiago, Chile
 }

\begin{document}

\date{Accepted ... Received ... in original form ...}


\maketitle

\label{firstpage}

\begin{abstract}
We present photometric and spectroscopic analysis of AE For -- a detached eclipsing binary
composed of two late K dwarfs. The masses of the components are found to be 0.6314$\pm$0.0035
and 0.6197$\pm$0.0034~$M_\odot$ and the radii to be 0.67$\pm$0.03 and 0.63$\pm$0.03~$R_\odot$
for primary and secondary component, respectively. Both components are significantly oversized
compared to theoretical models, which we attribute to their high activity. They show H$_\alpha$,
H$_\beta$, H$_\gamma$, Ca H and Ca K in emission, and are heavily spotted, causing large
variations of the light curve.

\end{abstract}


\section {Introduction}
 \label{sect: intro}

The 10th magnitude star AE For (CD -25 1273; HIP 14568) was first
classified by \citet{ste86} as a K4 dwarf. In the Hipparcos
catalogue \citep{esa97} it was listed as a new eclipsing variable of
Algol type with a period of 0.918235(8)~d, and a parallax of
32.10$\pm$1.78~mas. The latter was revised in 2007 to the presently
adopted 31.8$\pm$1.96~mas \citep{vle07}, corresponding to
31.5$\pm$1.9~pc.

\citet{giz02} found the star to be a double-lined binary with Balmer
lines in emission. Emission in Balmer and Ca II H \& K lines was
also observed by \citet{gra06}, who estimated the spectral type of
AE For at K9 Ve, and classified it as a very active system. A
slightly earlier spectral type (K7~Ve) was assigned to AE For by
\citet{tor06}, who also noted its high activity.
According to Zasche, Svoboda \& Uhl\'a\v{r} (2012; hereafter ZSU),
the system contains a brown dwarf with a minimal mass of $\sim$47 $M_{Jup}$ 
on an eccentric orbit of $\sim$7 years.

In the solar neighborhood about 7 per cent of detached eclipsing
binaries are X-ray emitters \citep{szcz08}. With its
$L_X=9.4\times10^{29}$~erg~s$^{-1}$ \citep{hue99} and
$F_X/F_{opt}=0.01$ \citep{fis99}, AE For is a prominent member of
this group. In the catalogue of \citet{szcz08} it is one of the closest 
X-ray active binaries, whose $L_X/L_{opt}$ ratio reaches 0.01 \citep{fis99}. 
An X-ray flaring activity of the system was reported by \citet{fuhr03}. 
It is also a bright IR source, with 2MASS magnitudes $J = 7.52$, $H = 6.85$ 
and $K = 6.66$. 

For about a decade it has been known that active dwarfs of K and M
type tend to be larger and cooler than the theory predicts. While
possible solutions of this problem have been proposed (see e.g.
Morales, Ribas \& Jordi 2008 and references therein), it is
certainly worthwhile to enlarge the relevant observational
databasis. The best opportunity for this is offered by detached
binaries on the lower main sequence, which allow to determine masses
and radii of their components with an accuracy better than 1 per
cent. AE For is clearly one of such systems, however until very
recently neither the light curve nor the velocity curve of this
potentially interesting binary has been studied. A preliminary
light-curve solution has been derived by ZSU, who concluded that a
spectroscopic analysis was needed to confirm the physical parameters
of the components to a higher accuracy.

In the present paper we obtain and analyze the velocity curve of AE
For, and refine the photometric solution of ZSU based on additional
observations. The photometric and spectroscopic data are described
in Sects. 2 and 3. The analysis of the data is detailed in Sect. 4,
and its results are discussed in Sect. 5.
\section {Photometric observations}
\label{sect:obsphot}

The observational material consists of our own $BV$ data and $BVR$
data of ZSU (see Table~\ref{tab:obsphot} for details). Our data
were collected during 11 nights
between 2009 Dec 8/9 and Dec 18/19 with the 1-m Swope telescope at
Las Campanas Observatory (LCO), Chile. All the nights were clear,
with the seeing between 1.1 and 2.5~arcsec. AE For was monitored
with the $2048\times3150$~pixel SITe3 CCD camera at a scale of
0.435~arcsec/pixel. We collected 482 frames through the $V$ filter with
exposure times 5-10~s (depending on the seeing), and 160 frames in
the $B$ filter with exposure times 8-25~s. All images were de-biased
and flat-fielded within the IRAF\footnote{IRAF is distributed by the
National Optical Astronomy
 Observatories, which are operated by the AURA, Inc., under cooperative
 agreement with the NSF.}

For photometric measurements the Daophot package \citep{ste87} was
employed, and differential aperture photometry was extracted using
two stars in the vicinity of AE~For. On every night AE~For was
observed we also observed NGC 2204. For six red clump stars of the
latter we found the offset between instrumental magnitudes and the
standard $BV$ magnitudes determined by \citet{roz07}. Using that
offset together with the average extinction for LCO 
\citep[$k_V = 0.14$ and $k_B = 0.24$; see][]{min89}, and accounting for 
a 0.04 difference in air-mass
between AE For and NGC 2204, we transformed the instrumental light
curves of the variable to the standard $BV$ system. Color terms of
the transformation were accounted for. We estimate the error of this
transformation at $\pm$0.02~mag, its main source being the
uncertainty of the zero-point for NGC 2204. The formal error of the
differential photometry is about ten times smaller.

\setlength{\tabcolsep}{5pt}
\begin{table}
  \caption{List of light curves used in this paper for photometric solutions of AE For.
  \label{tab:obsphot}}
   \begin{tabular}{ccccc}
    \hline
       1$^{\mathrm st}$ day  & last day  &  Filter & Number of  & Ref.  \\
       \multicolumn{2}{c}{[HJD-2400000]} &         & data points&       \\
    \hline
       55174& 55184&  Johnson $B$   & 160 & 1\\
       55174& 55184&  Johnson $V$   & 482 & 1\\
       55568& 55578&  Johnson $B$   & 173 & 2\\
       55568& 55578&  Johnson $V$   & 175 & 2\\
       55568& 55578&  Cousins $R$   & 173 & 2\\
    \hline
   \end{tabular}\\
\rule{0 mm}{4 mm} {\small 1: our own data; 2: \citet{zas12}.}
\end{table}

The last three rows in Table~\ref{tab:obsphot} refer to $BVR$ measurements
performed by ZSU in January 2011 at the South African Astronomical Observatory,
(their data are accessible
online at http://vizier.cfa.harvard.edu/viz-bin/VizieR?-source=J/A+A/537/A109).
Accessible online were also Hipparcos \citep{esa97}, ASAS \citep{poj02}, and Pi
of the Sky \citep{bur05} light curves. However, their quality was too poor to use
them for photometric solutions.

\begin{figure}
 \includegraphics[width=8.5cm,bb= 31 295 565 692,clip]{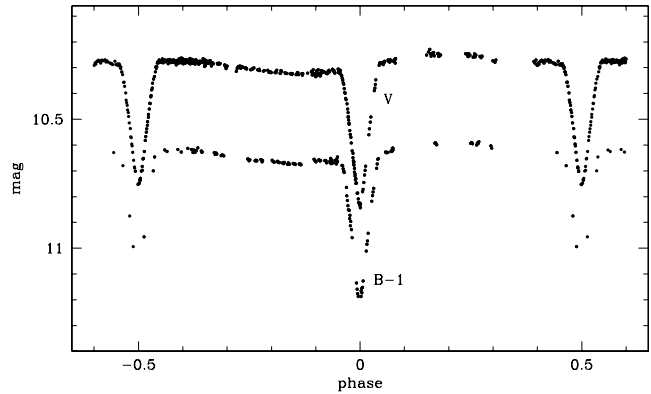}
 \caption {Light curves of AE For obtained in December 2009 at Las
           Campanas Observatory. To save the plotting space the 
           $B$-curve is shifted upward by one magnitude.
           \label{fig:PPlc}
          }
\end{figure}

Based on almost 70 primary and secondary minima observed by various authors,
ZSU derived the following ephemeris of AE For:
\begin{equation}
 HJD_\mathrm{min} = 2452605.97070(35) + 0.91820943(12)\times E\\
 \label{eq:ephem}\\
\end{equation}
The final light curves obtained from the data collected at LCO and
phased with this ephemeris are shown in
Fig.~\ref{fig:PPlc}. Evident is a strong asymmetry, which we
attribute to the activity of the system, manifesting itself by
large spot(s) on at least one of the components (see also Sects.
\ref{sect: obsspec} and \ref{sect:analysis}). For the short flat
part of the $V$-curve which begins right after the secondary eclipse
(see Fig. \ref{fig:PPlc}) we obtained $V = 10.27$$\pm$0.02~mag. To
within the errors, our $(B-V)$ index was constant throughout the
orbital period, and equal to 1.35$\pm$0.03~mag.
For the same part of the light curve the online data of ZSU yield $V = 10.23\pm0.01$~mag
and $(B-V)=1.34\pm0.02$~mag. Marginal differences between our results and theirs may be
due to the spot-related variability of at least one component of AE For.
\section {Spectroscopic observations}
\label{sect: obsspec} The spectroscopic data were collected with the
Echelle Spectrograph on the 2.5-m Ir\'en\'ee du Pont telescope at
LCO. The spectrograph, equipped with the SITe2K CCD camera, was
working at a slit width of 1.0 arcsec, providing a resolution of
$\sim$45,000. The observations were performed in two runs: four
nights from 2009 Nov 23/24 to Nov 26/27 and five nights from 2010
Dec 19/20 to Dec 23/24. All the nights were clear but a half of one
night with thin cirrus clouds. For most ($\sim$90 per cent)
exposures the airmass was smaller than 1.4. During the observations
pairs of 480-s exposures of the scientific target were made,
interlaced with a 90-s exposure of a ThAr lamp spectrum.

The observations were reduced within the IRAF ECHELLE package. After
bias and flat-field correction, each pair of the frames was combined
into a single frame, allowing for the rejection of cosmic ray hits.
Altogether, 31 reduced spectra were obtained, extending from
$\sim$4000 to $\sim$7000~\AA. All of them showed strong Balmer
emission lines, and in nearly all of them the NaI doublet at 5890
and 5896~\AA\ was blended into a very broad ($\sim$15~\AA, or
$\sim$760~km~s$^{-1}$) absorption feature.

Radial velocities were measured using an implementation of the
broadening function formalism \citep{Ruc02} described by
\citet{Kal06}.\footnote{The software package used in this paper is
freely accessible at http://users.camk.edu.pl/pych/BF/.} Since
including emission lines or broad absorption features would decrease
the accuracy of velocity measurements, a wavelength range devoid of
such features had to be selected. Shortward of H$_\gamma$ the
spectra were too noisy, so that our choice was reduced to four
segments limited by H$_\gamma$ H$_\beta$, NaI doublet, H$_\alpha$
and O$_2$ b-band. We chose the segment between H$_\beta$ and NaI
doublet, extending from 4870 to 5845~\AA. It was the longest one of
the four, and the only one in which the mean S/N ratio was larger
than 20 for all the spectra. The solar-scaled synthetic spectrum for
$T=4000$~K and $\log g=5.0$ from the library of \citet{Coe05} served
as a template.
\begin{table}
  \caption{Barycentric radial velocities of AE For phased according
           to the ephemeris given by equation (\ref{eq:ephem}).
           \label{tab:vc}}
   \begin{tabular}{rrrr}
    \hline
    HJD-2455000 & $v_{\rm p}$ [km s$^{-1}$]& $v_{\rm s}$ [km s$^{-1}$]& phase\\
    \hline
159.57213   &   9.88    &   110.40  &   0.06644 \\
159.63862   &   -28.91  &   148.95  &   0.13885 \\
159.70775   &   -52.93  &   177.94  &   0.21414 \\
159.77389   &   -53.68  &   175.07  &   0.28617 \\
159.78416   &   -51.58  &   173.19  &   0.29736 \\
160.52630   &   -14.47  &   134.61  &   0.10560 \\
160.62988   &   -53.98  &   175.83  &   0.21841 \\
160.64699   &   -56.77  &   179.20  &   0.23704 \\
160.67158   &   -57.13  &   179.58  &   0.26382 \\
160.75599   &   -30.58  &   151.97  &   0.35575 \\
161.61738   &   -52.80  &   174.11  &   0.29387 \\
161.73443   &   3.17    &   117.71  &   0.42135 \\
162.53278   &   -53.57  &   174.34  &   0.29081 \\
162.54966   &   -49.00  &   170.00  &   0.30920 \\
162.60757   &   -24.52  &   145.63  &   0.37227 \\
162.64129   &   -5.39   &   126.63  &   0.40899 \\
162.81241   &   126.34  &   -7.49   &   -0.40465\\
162.82922   &   138.43  &   -20.31  &   -0.38634\\
550.55538   &   143.22  &   -24.08  &   -0.12303\\
550.57283   &   129.29  &   -10.95  &   -0.10402\\
550.58045   &   124.95  &   -6.27   &   -0.09573\\
551.73601   &   -38.74  &   161.16  &   0.16277 \\
551.75427   &   -47.05  &   168.90  &   0.18265 \\
552.68821   &   -50.75  &   173.13  &   0.19979 \\
552.70336   &   -54.04  &   175.98  &   0.21629 \\
552.71858   &   -56.71  &   178.77  &   0.23286 \\
553.59737   &   -49.76  &   170.50  &   0.18993 \\
553.67500   &   -54.54  &   177.94  &   0.27448 \\
553.73276   &   -39.80  &   162.90  &   0.33738 \\
554.61505   &   -50.65  &   172.56  &   0.29827 \\
554.67243   &   -27.41  &   150.57  &   0.36076 \\
    \hline
   \end{tabular}
\end{table}

\begin{figure}
 \includegraphics[width=8.5cm,bb= 31 295 565 692,clip]{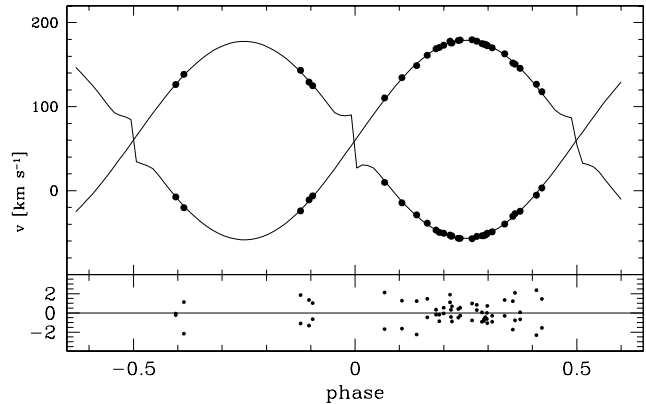}
 \caption {Velocity curve of AE For based on the data contained in Table
           \ref{tab:vc}. The rms residual velocity calculated from
           all (i.e. primary's and secondary's) residua shown in the
           bottom panel is equal to 1.13 km~s$^{-1}$.
           Phase~0 corresponds to the center of the primary photometric
           minimum.
           \label{fig:vc}
          }
\end{figure}

The observed velocity curve was fitted with a nonlinear
least-squares solution, using a spectroscopic data solver written
and kindly provided by Guillermo Torres. Because to within
observational errors secondary minima are separated from the primary
ones by half the period, a circular orbit was assumed, and the
eccentricity $e$ was fixed at 0 (the same assumption was adopted by
ZSU). The resulting barycentric radial velocities are listed in
Table~\ref{tab:vc} and plotted in the upper panel of
Fig.~\ref{fig:vc} together with the fitted velocity curve. The mean
error of velocity measurement, estimated from the residual
velocities shown in the bottom panel of Fig.~\ref{fig:vc}, is
$\pm$1.13~km~s$^{-1}$. Its rather large value may result from line
blending and/or asymmetries (the latter being caused by large spots;
see also Sects. \ref{sect:obsphot} and \ref{sect:analysis}). We
repeated the measurements using a synthetic spectrum for $T=4250$~K
to find that the changes in the observed
velocities were much smaller than the scatter of points around the
fit. We also tried out the remaining three wavelength ranges,
however in all cases the fit had a lower quality (i.e. the
resi\-dual velocities were larger).

The derived orbital parameters are listed in Table \ref{tab:orb_parm} 
together with formal 1-$\sigma$ errors returned by the fitting
routine. Note that, despite the rather large residual velocities,
the solution is fairly accurate - relative errors in $M\sin^3i$ and
$A\sin i$ amount to 0.5 and 0.15~per~cent, respectively.
%
%
\section {Light curve analysis and system parameters}
 \label{sect:analysis}

The analysis of the light curves was performed with the PHOEBE
implementation \citep{Prs05} of the Wilson-Devinney (WD) model
\citep{wil71,wil79}. The PHOEBE/WD package utilizes the Roche
geometry to approximate the shapes of the stars, uses Kurucz model
atmospheres, treats reflection effects in detail, and, most
importantly, allows for the simultaneous analysis of $B$ and $V$
data. Appropriately for stars with convective envelopes, we adopted
gravity darkening coefficients $g_p=g_s=0.32$ and bolometric albedos
$A_p=A_s=0.5$. The effects of reflection were included, and a
logarithmic limb-darkening based on tables by \cite{VHa93} was used
as implemented in PHOEBE 031a. Following ZSU, full synchronization
of both components was assumed.

ZSU fitted the light curves of AE For with the help of the program
ROCHE \citep{pri04}. Keeping in mind that their ROCHE and PHOEBE
fits of GK Boo were markedly different (although not at a
statistically significant level), we decided to check if our
photometric solution of AE For will be compatible with theirs. We
fed their photometric data into PHOEBE, and iterated upon
inclination $i$, secondary's temperature $T_s$, primary's radius
$R_p$ and secondary's radius $R_s$, using values obtained by ZSU as
initial values. The mass ratio $q$ was fixed at a value of 0.98522
calculated from Table~\ref{tab:orb_parm}, and the orbital separation
$A$ -- at 4.28996~$R_\odot$ obtained from $A\sin i $ in
Table~\ref{tab:orb_parm} for the inclination $i=85.51^\circ$ found
by ZSU. Following ZSU, we kept the temperature of the primary $T_p$
fixed at 4100~K. To within the errors, this value agrees with
4055$\pm$71~K resulting from the $T_{eff}-(B-V)$ calibration of
\citet{ram05}, and marginally agrees with 4005$\pm$92~K resulting
from the analogous calibration of \citet{wor11}. The latter two values, however,
should be treated with some caution, since AE For emits most energy
in the near-IR range beyond $B$ and $V$ passbands, at 2MASS
magnitudes $J = 7.52$, $H = 6.85$ and $K = 6.66$. An infrared
calibration would be more appropriate; unfortunately we have not
found any such relation extending up to $(J-K)=0.86$. The problem of
the temperature is further discussed in Sect. \ref{sect:discussion}.
\begin{table}
 \caption{Orbital parameters
          \label{tab:orb_parm}
         }
 \begin{tabular}{lcc}
  \hline
   Parameter  & Unit &  Value \\
  \hline
     $\gamma$   &km s$^{-1}$ & 59.95$\pm$0.23 \\
     $K_p$      &km s$^{-1}$ &116.92$\pm$0.33 \\
     $K_s$      &km s$^{-1}$ &119.12$\pm$0.33 \\
     $e$        &     & 0.0$^\mathrm{a}$ \\
     $\sigma_p$   &km s$^{-1}$ & 1.08    \\
     $\sigma_s$   &km s$^{-1}$ & 1.18    \\
     Derived quantities:        &                 &\\
     $A\sin i$    &$R_\odot$ & 4.2820$\pm$0.0069 \\
     $M_p\sin^3 i$  &$M_\odot$ & 0.6314$\pm$0.0034 \\
     $M_s\sin^3 i$  &$M_\odot$ & 0.6197$\pm$0.0033 \\
  \hline
 \end{tabular}\\
\rule{0 mm}{4 mm}$^\mathrm{a}$Assumed in fit
\end{table}

\begin{figure}
 \includegraphics[width=8.5cm,bb= 19 362 565 692,clip]{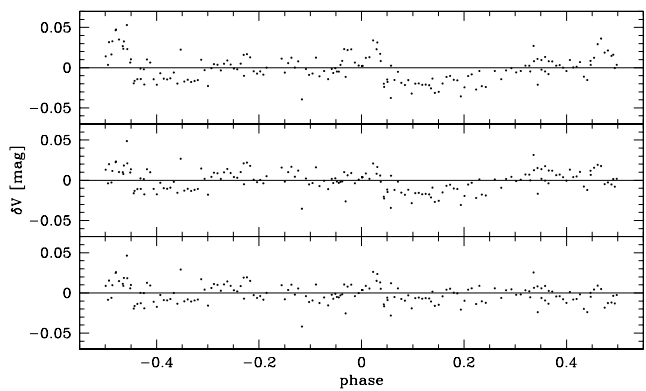}
 \caption {Residuals of fits listed in columns 3--5 of Table \ref{tab:phot_parm_ZSU}.
           Top: original ZSU: middle: PHOEBE-iterated starting from ZSU parameters,
           bottom: same as middle, but with spots added on both components
           (see text for details).
           \label{fig:zsu_res}
          }
\end{figure}

The results of fitting are listed in columns 3 and 4 of Table
\ref{tab:phot_parm_ZSU}. Column 3 contains the original values
obtained by ZSU together with their errors. The corresponding
synthetic light curves produced by PHOEBE without any iterations
turned out to be fairly well fitting, with a standard deviation of
$V$-residuals equal to 17~mmag (similar values were obtained for $B$
and $R$ bands). The iterations improved the quality of the fit
($\sigma_V$ was reduced to 13~mmag), but the parameters did not
change in a statistically significant way (column 4). We performed 
an additional iteration with the following spots introduced on both 
components at a latitude of 90$^\circ$ (i.e. at the equator in PHOEBE's 
convention): one on the primary at a longitude of 180$^\circ$ with a
radius of 20$^\circ$ and temperature factor of 0.98; two on the
secondary at longitudes = 125$^\circ$ and 330$^\circ$, both of them
with a radius of 30$^\circ$ and temperature factor of 1.02. The 
residuals became more symmetric (see Fig. \ref{fig:zsu_res}) and 
$\sigma_V$ was reduced to 12~mmag, but no statistically significant
corrections to system parameters were obtained (column 5 of Table
\ref{tab:phot_parm_ZSU}). We note that by ``spot'' we mean a fairly 
large region on the surface of a star, whose mean temperature is 
elevated or reduced due to an excess of genuine stellar spots with much 
smaller sizes.
\noindent
\begin{table}
 \caption{Photometric parameters derived from ZSU data. The lower
          bounds for component masses derived from the orbital solutions
          are 0.6314$\pm$0.0034 and 0.6197$\pm$0.0033~$M_\odot$, respectively,
          for the primary and the secondary.
          \label{tab:phot_parm_ZSU}
         }
 \begin{tabular}{lccccc}
  \hline
   Parameter      & Unit &  Original       &  Our fit & Our fit     & Dark\\
                  &      &  ZSU            &          & with spots  & pole\\
       1          &  2   &   3             &     4    &    5        &  6  \\
  \hline
     $i$          & deg  & 86.51(31) & 86.84 & 86.71 & 85.47\\
     $T_s$        & K    & 4065(48)  & 4083  & 4052  & 4041 \\
     $R_p$        & $R_\odot$ & 0.66(10) & 0.71 & 0.71 & 0.661\\
     $R_s$        & $R_\odot$ & 0.52(8)  & 0.53 & 0.52 & 0.605\\
     $(L_p)_V$    & \%   & 63.1(1.2) & 65.4 & 66.8 & 57.2\\
     $(L_p)_B$    & \%   & 63.2(1.3) & 65.6 & 67.3 & 57.9\\
     $\sigma_V$ &mmag& 17 & 13 & 12 & 13 \\
  \hline
 \end{tabular}
\end{table}

For the assumed $M_s = 0.5M_\odot$, ZSU derived $R_s = 0.52R_\odot$,
and our PHOEBE iterations confirmed this result. However, while a
radius of $0.52R_\odot$ may be compatible with the mass of $M_s =
0.5M_\odot$ assumed by ZSU, it is excluded by our spectroscopic solution, in which
$M_s>0.6M_\odot$. This is because within a broad range of
metallicity ($-0.5\leq\mathrm{[Fe/H]}\leq0.5$) and
$\alpha$-enhancement ($-0.2\leq[\alpha\mathrm{/Fe]}\leq0.4$) a
main-sequence star with $M$=$0.6M_\odot$ never becomes that small
\citep[see][and the DSED evolutionary tracks at
http://stellar.dartmouth.edu/models/index.htm]{Dot08}. Another
argument in favor of a larger secondary is based on rotational
velocity fits which are performed automatically within the BF
formalism during radial velocity measurements. Upon averaging
velocities fitted in 2009 and 2010 we got $v_{\mathrm{rot},s}$ =
37.46$\pm$2.40~km~s$^{-1}$ for the primary and $v_{\mathrm{rot},s}$
= 37.07$\pm$2.20~km~s$^{-1}$ for the secondary. For the assumed
synchronous rotation these values imply $R_p= 0.680\pm0.044 R_\odot$
and $R_s = 0.673\pm0.040 R_\odot$.

The simplest way to remove this contradiction is to force a smaller
inclination $i$: one may expect that $R_s$ will then increase, so 
that depths and widths of the minima are preserved. This may
be achieved, for example, by placing a dark spot around the
uneclipsed pole of the primary. We performed an experiment, in which
the circumpolar region of the primary with a radius of 50$^\circ$
had a temperature factor of 0.95. As shown in column 6 of
Table~\ref{tab:phot_parm_ZSU}, the iterations yielded a solution
with $i$ smaller by more than one degree and $R_s$ larger by almost
0.1~$R_\odot$ compared to the model without a darkened pole.

At this point we faced the problem whether to include clearly 
unphysical solutions in final estimates of system parameters, and 
we decided to reject them. Apparently, ZSU collected their data when
the system was in a particular state making the analysis very uncertain,
and to remain on the safe side we calculated the parameters of AE For 
based on our photometry only.
\begin{table}
 \caption{Photometric parameters derived from our data
          \label{tab:phot_our_data}
         }
 \setlength{\tabcolsep}{4pt}
 \begin{tabular}{lcccccc}
  \hline
   Parameter      & Unit & P0  & P0    & P1h  & P1c   & Mean\\
                  &      & S0  & S1h2c & S2c  & S1h1c & col. 4-6 \\
       1          &  2   &   3 &   4   &   5  &  6    &  7    \\
  \hline
     $i$          & deg       & 85.72 & 85.75 & 85.51 & 85.43&85.6$\pm$0.2\\
     $T_s$        & K         & 4016  & 4055  & 4061  & 4050 &4055$\pm$6\\
     $R_p$        & $R_\odot$ & 0.705 & 0.694 & 0.665 & 0.650&0.67$\pm$0.03\\
     $R_s$        & $R_\odot$ & 0.603 & 0.607 & 0.640 & 0.654&0.63$\pm$0.03\\
     $(L_p)_V$    & \%        & 61.6  & 58.8  & 53.8  & 52.1 &54.9$\pm$4.4\\
     $(L_p)_B$    & \%        & 62.5  & 59.4  & 54.2  & 52.7 &55.4$\pm$4.4\\
     $\sigma_V$   &mmag       & 20.1  &  7.0  & 6.5   & 6.5  &--\\
  \hline
 \end{tabular}
\end{table}
\noindent
\begin{table}
 \caption{Parameters of spots for models from Table \ref{tab:phot_our_data}
          \label{tab:spot_parm}
         }
 \begin{tabular}{lcccc}
  \hline
   Model      & component &latitude      & radius & temp.factor \\
  \hline
   P0S1h2c    & S       &  60 & 40 & 0.99 \\
              & S       & 100 & 30 & 1.02 \\
              & S       & 230 & 30 & 0.95 \\
   P1hS2c     & P       & 280 & 30 & 1.02 \\
              & S       &  60 & 40 & 0.99 \\
              & S       & 230 & 30 & 0.95 \\
   P1cS1h1c   & P       & 240 & 40 & 0.99 \\
              & S       & 100 & 30 & 1.02 \\
              & S       & 230 & 30 & 0.95 \\
  \hline
 \end{tabular}
\end{table}

\begin{figure}
 \includegraphics[width=8.5cm,bb= 19 459 565 692,clip]{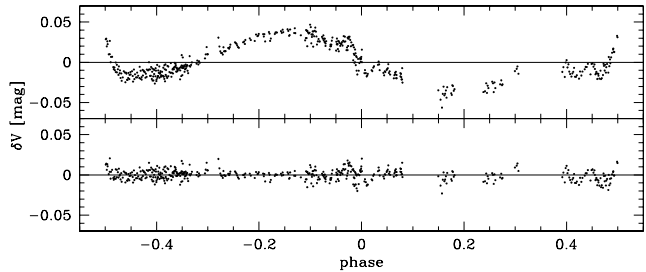}
 \caption {Residuals of fits listed in columns 3 and 4 of Table \ref{tab:phot_our_data}.
           Top: model P0S0 (without spots); bottom: model P0S1h2c (three spots on the
           secondary; see Table \ref{tab:spot_parm} for details).
           \label{fig:pp_res}
          }
\end{figure}

Photometric solutions based on our own data are listed in Table
\ref{tab:phot_our_data}. Four models are shown, identified by the
numbers of hot (h) and cold (c) spots on the primary (P) and the
secondary (S). In each spotted model there are three spots centered
at a latitude of 90$^\circ$. The remaining parameters of the spots
are listed in Table \ref{tab:spot_parm}. The fit without spots
(P0S0) is poor, as indicated by a large $\sigma_V$. Its low quality
is clearly seen in Fig. \ref{fig:pp_res}, in which the residuals
show systematic deviations with an amplitude exceeding 40 mmag.
Introducing spots makes the fit nearly ideal. Not surprisingly,
however, Table \ref{tab:phot_our_data} shows that different spot
arrangements result in different system parameters. The encouraging
finding is that for all models (even for the unspotted one) the
inclination angle is by about one degree smaller than that obtained
from ZSU data, and the radius of the secondary is consistently
larger than 0.6$R_\odot$. The last column of
Table~\ref{tab:phot_our_data} contains averaged values of spotted
model parameters together with formally calculated errors. Since the
distribution of spot-dependent model parameters is most probably 
non-Gaussian, these errors should be treated with caution, and we 
give them only to approximately illustrate the uncertainties resulting 
from the nonexistence of a unique configuration of spots.

Independently of their spot-related uncertainty, the models have
Gaussian errors related to nonzero residuals, which we estimated
using a Monte Carlo procedure written in PHOEBE-scripter. Briefly,
the procedure replaces the observed light curves $B_{obs}$ and
$V_{obs}$ with the fitted ones $B_f$ and $V_f$, generates 20000
Gaussian perturbations $\delta B_f$ and $\delta V_f$ such that the
standard deviation of each of them is equal to the standard
deviation of the residuals $B_f-B_{obs}$ or $V_f-V_{obs}$, and for
each perturbation performs PHOEBE iterations on $V_f+\delta V_f$ and
$B_f+\delta B_f$. In all cases the Gaussian errors turned out to be
much smaller than the formal errors of averaged parameters given in
column 7 of Table~\ref{tab:phot_our_data}.

The final absolute parameters of AE For are given in Table
\ref{tab:abs_parm} (bolometric magnitudes in the last two rows are
averages of the values taken directly from PHOEBE output). The
errors of $A$, $M_p$ and $M_s$ include the inclination uncertainty.
As we explained above, the errors of the remaining parameters are
but an approximate illustration of the uncertainties related to the
presence of spots.
\noindent

\section {Discussion and conclusions}
 \label{sect:discussion}

The fits described in Sect. \ref{sect:analysis}  indicate that an
unavoidable consequence of spot activity is the degradation of the
photometric solution: based on the available data, the radii of the
components cannot be calculated with an accuracy better than
$\sim$5~per~cent. As we have shown in Sect. \ref{sect:analysis}, the 
asymmetric and highly variable light curve of AE For can even generate 
nonphysical solutions in which one of the stars is unrealistically 
small.  The only way to
improve the accuracy of system parameters is via collecting more
photometric data from various seasons and careful fitting of each
light curve with various arrangements of spots \citep[see
e.g.][]{roz09}. Alternatively, many light curves from various
seasons might be averaged in hope that the mean curve would be free
from spot-caused irregularities.

\begin{figure}
 \includegraphics[width=8.5cm,bb= 22 360 565 692,clip]{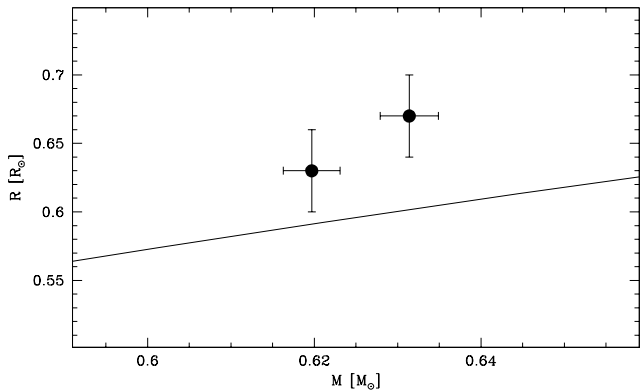}
 \caption {Location of our final solution in the $(M-R)$ plane. Solid 
           line: main sequence represented by the 1~Gyr Dartmouth 
           isochrone for solar abundances. 
           \label{fig:mr_plane}
          }
\end{figure}

Another potential source of uncertainties in our final
parameters is the effective temperature of the primary, which we
fixed at 4100~K (i.e. a value adopted by ZSU). Luckily, the temperature only
weakly affects $i$, $R_s$ and $R_p$ -- we checked that they all
change by less than 0.2 per cent for a $\pm$150~K change in $T_p$.
It is thus possible to compare the derived radii with stellar
models, and such a comparison with the synthetic main sequence 
represented by the 1~Gyr Dartmouth isochrone for [Fe/H] =
0.0 and [$\alpha$/Fe] = 0.0 is shown in Fig. \ref{fig:mr_plane}.
Each component of AE For is clearly oversized vs. a main-sequence
star of the same mass. This conclusion holds for a broad range of chemical
composition parameters ($-0.5\leq\mathrm{[Fe/H]}\leq0.5$;
$-0.2\leq[\alpha\mathrm{/Fe]}\leq0.4$). 
\begin{table}
 \caption{Absolute parameters of AE For
          \label{tab:abs_parm}
         }
 \begin{tabular}{lcc}
  \hline
   Parameter     & Unit & Value  \\
  \hline
     $A$   & $R_\odot$  & 4.2820$\pm$0.0069 \\
     $i$   &  deg       &  85.6 $\pm$0.2$^\mathrm{a}$    \\
     $e$   &            &   0$^\mathrm{b}$  \\
     $M_p$ & $M_\odot$  & 0.6314$\pm$0.0035 \\
     $M_s$ & $M_\odot$  & 0.6197$\pm$0.0034 \\
     $R_p$ & $R_\odot$  & 0.67$\pm$0.03$^\mathrm{a}$     \\
     $R_s$ & $R_\odot$  & 0.63$\pm$0.03$^\mathrm{a}$     \\
     $T_p$ & K          & 4100$^\mathrm{b}$ \\
     $T_s$ & K          & 4055$\pm$6$^\mathrm{a}$        \\
     $M^{bol}_p$ & mag  &7.15$\pm$0.09$^\mathrm{a}$ \\
     $M^{bol}_s$ & mag  &7.32$\pm$0.11$^\mathrm{a}$ \\
  \hline
 \end{tabular}\\
\rule{0 mm}{3 mm}
$^\mathrm{a}$Approximate errors due to spots (see text)\\
\rule{1 mm}{0 mm}$^\mathrm{b}$Assumed in fit
\end{table}

\begin{figure}
 \includegraphics[width=8.5cm,bb= 22 360 565 692,clip]{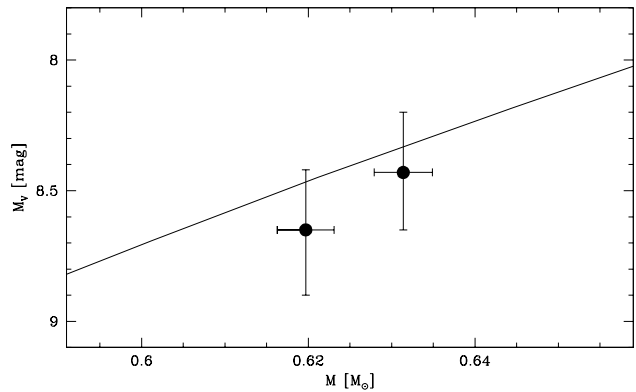}
 \caption {Location of AE For components in the $(M-M_V)$ plane.
           The magnitudes are derived from the observational data.
           Solid line: the same isochrone as in Fig. \ref{fig:mr_plane}.
           \label{fig:obsHR}
          }
\end{figure}

While active components of eclipsing binaries appear to be larger 
and cooler than inactive single stars of the same mass, they have 
a similar luminosity \citep{mor08}. To check if this holds for our binary,
we calculated absolute magnitudes of the components in $V$-band from 
the observed magnitude and parallax of the system, using the contribution 
of the primary to the total light from Table \ref{tab:abs_parm}. The 
location of the components in the $(M-M_V)$ plane is shown in 
Fig.~\ref{fig:obsHR} together with the main sequence represented 
as before by the 1 Gyr Dartmouth isochrone for solar abundances. As one
can see, to within the errors both stars do indeed belong to the main 
sequence. In the following we assume that they have main-sequence 
luminosities, and based on that assumption we estimate their temperatures.

\begin{figure}
 \includegraphics[width=8.5cm,bb= 22 360 565 692,clip]{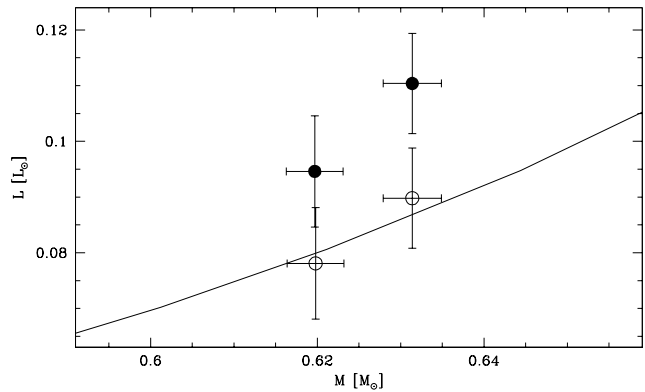}
 \caption {Location of our solutions in the $(M-L)$ plane. 
           Models with $T_p = 4100$ and $T_P=3900$ K are identified 
           by filled and open circles, respectively.
           Solid line: the same isochrone as in Fig. \ref{fig:mr_plane}.
           \label{fig:ml_plane}
          }
\end{figure}

Fig.~\ref{fig:ml_plane} indicates that the solution with $T_p=4100$~K is 
significantly too luminous. Only at $T_p=$ 3900~K (for which $T_s=3860$~K) 
do both stars align with the main sequence, suggesting that the temperature 
assumed by ZSU and the temperatures derived from the calibrations of
\citet{ram05} and \citet{wor11} are overestimated. Of course, this
finding should be verified, preferably by using disentangling
software \citep[see e.g.][]{had09}. Further spectroscopic
observations are needed for that purpose, collected during
low-activity periods. When the system is active, all temperature
estimates, whether based on spectral fitting or photometric
calibrations, are likely to be flawed because of large spotted areas
present on both components. 

We have thus proved the consistency of the assumption that the components 
of AE For are main-sequence stars. However, given the large errors of $M_V$,
a possibility that they have not yet settled on the main sequence should
also be explored. If they were indeed in the contraction phase, their large 
sizes could be at least partly caused by evolutionary effects. Such a possibility 
is indicated by the Li~6708~\AA\ line identified in the spectrum of AE For by 
\citet{tor06}. With an equivalent width of 80~m\AA\ it falls right in the middle 
of the range observed for Pleiades members with the same $(V-I)$, suggesting that 
the system is rather young. 
Unfortunately, our efforts to repeat the measurement of \citet{tor06} were 
unsuccessful. We do not claim that the line is not there: our spectra may be too 
noisy, or the broadening is too strong, or both. It is worth mentioning, however,
that based on kinematic criteria, \citet{egg90} assigned AE~For to the Hyades 
supercluster whose age is estimated at $\sim$0.6~Gyr \citep{mon01}. If this 
assignment is correct, then both components of the binary must have ended their 
pre-main sequence evolutionary phase long ago. Again, high-quality spectra, 
collected preferably during eclipses, are needed to resolve the Lithium (and age) 
problem. 

Yet another explanation for the oversized components involves the 
third body discovered by ZSU. In a triple system consisting of a binary orbited 
by a distant companion both the eccentricity of the binary $e_b$ and the 
inclination of the third body's orbit execute periodic Kozai oscillations. If at 
some phase of the oscillation cycle $e_b$ becomes sufficiently large for tidal 
friction to dissipate the orbital energy of the binary, then the binary  
gradually tightens its orbit \citep[see e.g.][and references therein]{fab07}. 
The associated tidal heating could temporarily inflate the components; however,
since nearly all close binaries are members of triple systems \citep{tok06}, and 
the inflation effect is limited to low-mass stars of K and M type \citep{mor08}, 
we consider this explanation much less likely than the remaining two (especially
than that related to stellar activity).  
  
The above discussion indicates that AE For is a truly interesting object
which clearly deserves closer attention. 
A systematic study of this system would certainly bring valuable information
concerning the activity of late-type stars.

\section*{Acknowledgments}
We are greatful to the anonymous referee for the detailed and helpful 
report.
Support for R.A. is provided by Proyecto GEMINI CONICYT \#32100022 and 
via a Postdoctoral Fellowship by the School of Engineering at Pontificia 
Universidad Cat\'olica de Chile.
Support for I.D. is provided by the Chilean Ministry for the Economy, 
Development, and Tourism’s Programa Inicativa Cient\'{\i}fica Milenio through 
grant P07-021-F, awarded to The Milky Way Millennium Nucleus, and by Proyecto 
FONDECYT Regular \#1110326.
We thank Guillermo Torres for providing the spectroscopic data solver.
This research has made use of the SIMBAD database, operated at CDS,
Strasbourg, France.

\clearpage

\section* {ONLINE MATERIAL}

\begin{table}
  \caption{A sample of the $B$-lightcurve of AE For.}
   \begin{tabular}{lr}
    \hline
       HJD-240000 & $B$ [mag] \\
    \hline
55174.610258 & 11.629\\
55174.629343 & 11.680\\
55174.650730 & 11.994\\
55174.673195 & 11.956\\
55174.695058 & 11.646\\
55174.715983 & 11.619\\
    \hline
   \end{tabular}\\
\end{table}

\begin{table}
  \caption{A sample of the $V$-lightcurve of AE For.}
   \begin{tabular}{lr}
    \hline
       HJD-240000 & $V$ [mag] \\
    \hline
55174.662443 & 10.745\\
55175.581451 & 10.744\\
55174.664191 & 10.731\\
55175.583256 & 10.733\\
55174.665765 & 10.725\\
55174.667675 & 10.697\\
    \hline
   \end{tabular}\\
\end{table}

\end{document}